# Chiral emergence in multistep hierarchical assembly of achiral conjugated polymers

Kyung Sun Park[1], Zhengyuan Xue[1], Bijal B. Patel[1], Hyosung An[2], Justin J. Kwok[2], Prapti Kafle[1], Qian Chen[2], Diwakar Shukla[1] and Ying Diao[1,2,3]*

[1]Department of Chemical and Biomolecular Engineering, University of Illinois at Urbana-Champaign, 600 S. Mathews Ave., Urbana, IL 61801, USA.

[2]Department of Materials Science and Engineering, University of Illinois at Urbana-Champaign, 1304 W. Green St., Urbana, IL 61801, USA.

[3]Beckman Institute, Molecular Science and Engineering, University of Illinois at Urbana-Champaign, 405 N. Mathews Ave., Urbana, IL 61801, USA.

*Corresponding author. Email: yingdiao@illinois.edu

**Abstract**

Intimately connected to the rule of life, chirality remains a long-time fascination in biology, chemistry, physics and materials science. Chiral structures, e.g., nucleic acid and cholesteric phase developed from chiral molecules are common in nature and synthetic soft materials. While it was recently discovered that achiral but bent core mesogens can also form chiral helices, the assembly of chiral microstructures from achiral polymers has rarely been explored. Here, we reveal chiral emergence from achiral conjugated polymers for the first time, in which hierarchical helical structures are developed through a multistep assembly pathway. Upon increasing concentration beyond a threshold volume fraction, pre-aggregated polymer nanofibers form lyotropic liquid crystalline (LC) mesophases with complex, chiral morphologies. Combining imaging, X-ray and spectroscopy techniques with molecular simulations, we demonstrate that this structural evolution arises from torsional polymer molecules which induce multiscale helical assembly, progressing from nano- to micron scale helical structures as the solution concentration increases. This study unveils a previously unknown complex state of matter for conjugated polymers that can pave way to a new field of chiral (opto)electronics. We anticipate that hierarchical chiral helical structures can profoundly impact how conjugated polymers interact with light, transport charges, and transduce signals from biomolecular interactions and even give rise to properties unimagined before.

**Main**

Hierarchical structures are inherent to various soft material systems including biomolecules, mesogens and conjugated polymers. Structural chirality commonly results from the hierarchical organization of chiral building blocks, e.g., amyloids, M13 phage and chiral mesogens[1-3]. In fact, chirality is common in nature and plays an essential role in various research fields such as biology, medicine, chemistry, physics and materials science[4-6]. For example, many biological substances including carbohydrates, amino acids and nucleic acids are chiral and their functional structures selectively respond to a particular chirality. Thus, understanding and controlling chirality across various length scales are important for developing functional soft materials. Recently, it has been discovered that achiral mesogens or metal oxide nanocubes can also form chiral, twisted structures due to symmetry breaking or topological defects[7, 8]. In particular, achiral bent shaped molecules such as rigid bent-core mesogens and dimers linked with flexible chains have exhibited chiral/helical assembled structures, e.g., helical nanofilament phases and twist-bend nematic phases[9-11].

When applied to semiconducting and conducting polymers, such achiral-to-chiral transitions open a new degree of freedom for tuning electrical, optical, biological and mechanical properties and can provide further fundamental understanding of complex supramolecular assembly and phase transition behaviors that occur during device fabrication. Conjugated polymers underpin a broad range of emerging technologies, due to their ability to transport charges[12, 13], couple light absorption with charge generation[12-14], couple ion transport with electron transport[15], and transduce various interactions and stresses into electrical signals[16, 17]. Thanks to such versatile functional properties, conjugated polymers have found use in electronics[18], thermoelectrics[19], solar cells[14], photocatalysts[20], electrochemical devices (fuel cells[21], batteries[22], supercapacitors[23]) and biomedical devices[17, 20]. All the fundamental physical processes described above sensitively depend on multiscale morphology that includes polymer conformation, packing, crystallinity, alignment and domain connectivity[24]. Such complex morphology is, in turn, highly sensitive to molecular assembly pathways[25, 26]. Particularly, recent studies have demonstrated that certain donor-acceptor (D-A) conjugated polymers readily form pre-aggregates and/or lyotropic liquid crystalline (LC) phases in appropriate solvent systems[27-33]. Such intermediate states have shown benefits of their molecular assembly in terms of increasing crystalline domain size, inducing alignment, and thus enhancing device performance of thin films. While the lyotropic LC phases of conjugated polymers have been observed, little is known of their structures and hierarchical assembly, let alone supramolecular chirality. Lack of critical structural information severely hinders our understanding of how LC phases of conjugated polymers impact thin film morphology and functional properties. Such knowledge on structure and in particular chiral emergence can further lead to novel optical, electronic, and mechanical properties unimagined before.

Herein, we report chiral emergence from achiral D-A conjugated polymers for the first time. We discover four liquid crystalline mesophases previously unknown to conjugated polymers, out of which three are chiral. We unveil a surprisingly complex hierarchical helical structures from the molecular, nanoscopic to micron scale. This insight is obtained by combining optical and electron microscopy imaging, optical spectroscopy, X-ray scattering with molecular dynamic (MD) simulations. Our findings are significant as the new states of matter discovered can redefine how we understand their optical, electronic, mechanical and biological properties of conjugated polymers which underpin a wide range of emerging technologies. This work also contributes to fundamental understanding of liquid crystals as this is the first report of polymeric twist-bent mesophases whose formation mechanism is distinct from bent-core molecules and colloids.

**Results and Discussion**

The D-A copolymer used in this study is an isoindigo-bithiophene-based copolymer (PII-2T) (Fig. 1a) which is slightly twisted caused by the systematic torsion within each repeat unit[33]. A detailed molecular structure in the solution state was further explored using MD simulations, discussed later. Such isoindigo-bithiophene-based copolymers belong to a family of high performance p-type organic semiconductors, extensively studied over the past 10 years as active materials for transistors, solar cells, bioimaging, etc[34]. However, no thermotropic nor lyotropic liquid crystals have been reported for isoindigo based conjugated polymers so far. We surprisingly discovered several new LC mesophases through cross-polarized optical microscopy (CPOM) and circular dichroism spectroscopy (CD). We then determined their nano- to micron-scale structures via electron microscopy techniques which revealed that the fundamental building blocks of the observed LC mesophases are nanofibril-like polymer pre-aggregates. The internal structures of the pre-aggregates were further analyzed via small-angle X-ray scattering (SAXS), grazing incidence wide angle X-ray scattering (GIWAXS) and UV-Vis absorption spectroscopy. This wide range of characterizations together with MD simulation leads us to depict on how complex hierarchical helical structures emerge from achiral conjugated polymers. These results are discussed in detail below.

**Discovery and identification of new mesophases.** We first observed emergence of several new liquid crystal phases from isotropic solution during evaporative assembly of PII-2T in a receding meniscus (Supplementary Information, Movie S1). Resembling the solution printing process, polymer solution traverses the entire concentration range from that of isotropic solution to that of solid thin film in the meniscus region. With increasing solution concentration, we observed nucleation of liquid crystals in confined droplets, merging of droplets into a continuous liquid crystal phase of uniform texture, followed by its transition to striped texture.

To investigate the structure of each phase in depth, a series of PII-2T solutions at defined concentrations was prepared by a drop-and-dry method from chlorobenzene solution. By this approach, successive drop-casting from a stock solution (10 mg/ml) was performed between two glass slides to concentrate solutions to above 200 mg/ml (see Materials and Methods). We note that, while chlorobenzene was used as the solvent, the same liquid crystal phases were observed in chloroform solutions as well. The morphology and optical properties of the PII-2T solutions are summarized in Fig. 1b, showing clearly that the mesophases are highly dependent on solution concentration, indicating a lyotropic LC nature. The solution prepared up to ~40 mg/ml displays no birefringence or apparent aggregates under CPOM. At ~50 mg/ml it begins to produce spindle-shaped birefringent microdroplets (tactoids), indicating a transition state where LC mesophases nucleate and grow from an isotropic phase. Detailed examination reveals that the mesophase at 50 mg/mL are, in fact, homogeneous nematic tactoids in which the director field, i.e., an average direction of the backbones is aligned to the long axis of the tactoids (Supplementary Information Fig. S1). As the solution concentration slightly increases to ~60 mg/ml, these homogeneous tactoids transition to bipolar tactoids, where the director field traces the curving edge of tactoids instead (Supplementary Information Fig. S2). The observed homogeneous to bipolar tactoid transition mirrors that of LC phases of amyloid fibrils when a specific critical volume is reached[1]. The tactoids observed in our study seem nonchiral as they show a mirror-symmetric birefringence with a dark center when the director is aligned parallel to either the polarizer or the analyzer (see Fig. S1 and S2), in contrast to optically active centers in the case of chiral tactoids[35]. This is because the twisted director in chiral tactoids forms a titled angle along the main axis of the tactoids. We note that this is only valid when the director is one handedness with a micron scale systematic twist. Further analysis of these nonchiral tactoids was carried out using CD discussed later. As the solution concentration further increases (~100 mg/ml), a uniform birefringent texture was observed, where discrete domains of isotropic phases (negative tactoids) are embedded in a continuous domain of mesophase. The solution at ~140 mg/ml showed complex morphology with micron-scale domains. At an extremely high concentration of ~200 mg/ml, a striped texture with a few micron periodicity was obtained, which resembles the zigzag twinned morphology of printed PII-2T solid films[33]. By rotating the sample under fixed crossed polarizers, the alternating dark and bright band patterns indicate the characteristic of twinned domains. Hereafter, we refer to each phase shown in Fig. 1b as (i) isotropic phase, (ii) nematic tactoids, (iii) twist-bent mesophase I, (iv) twist-bent mesophase II and (v) striped twist-bent mesophase that are proposed based on comprehensive structural characterizations, discussed later.

We next used CD spectroscopy to explore the chirality of the observed mesophases. Figure 1c shows the CD spectra of the PII-2T solutions at solution concentrations corresponding to the CPOM images shown in Fig. 1b. The crystalline mesophases made at 100 mg/ml and above show very intense bisignate CD bands near 630 nm which corresponds to the main absorption wavelength of the polymer solution (see Fig. 4a). This bisignate CD band represents chiral exciton coupling via Davydov splitting, indicating the polymer backbones are formed in a chiral fashion[36]. The mesophases at 100 and 140 mg/ml show negative and positive CD signals around 740 nm and 550 nm, respectively, indicating that the backbones form left-handed helical aggregation. The mesophase at >200 mg/ml shows inversed CD signs at each wavelength, indicating right-handed helical aggregation. This handedness inversion emerged at the same concentration corresponding to the formation of the twinned morphology. The origin of this handedness inversion is currently unknown, but some indication may be provided by comparison with natural cholesteric phases, where similar phenomena is often observed when the volume fraction (concentration) is high[37-39]. In these systems, the handedness inversion occurred when preferred packing structures resulted from a complex interplay of electrostatics, molecular sequences, excluded volume interactions, etc. Figure 1d shows a summary of the concentration normalized CD signals for various solution concentrations. In contrast to the twisted mesophases, the isotropic phase (<40 mg/ml) and the nematic tactoids (50-60 mg/ml) show zero or negligible CD signal. It is inferred that those phases are either nonchiral or form a near-racemic mixture. We speculate that the isotropic phase may exhibit both handedness equally (see discussions of MD results), while the nucleation and growth of mesophases drives the transition to single-handedness.

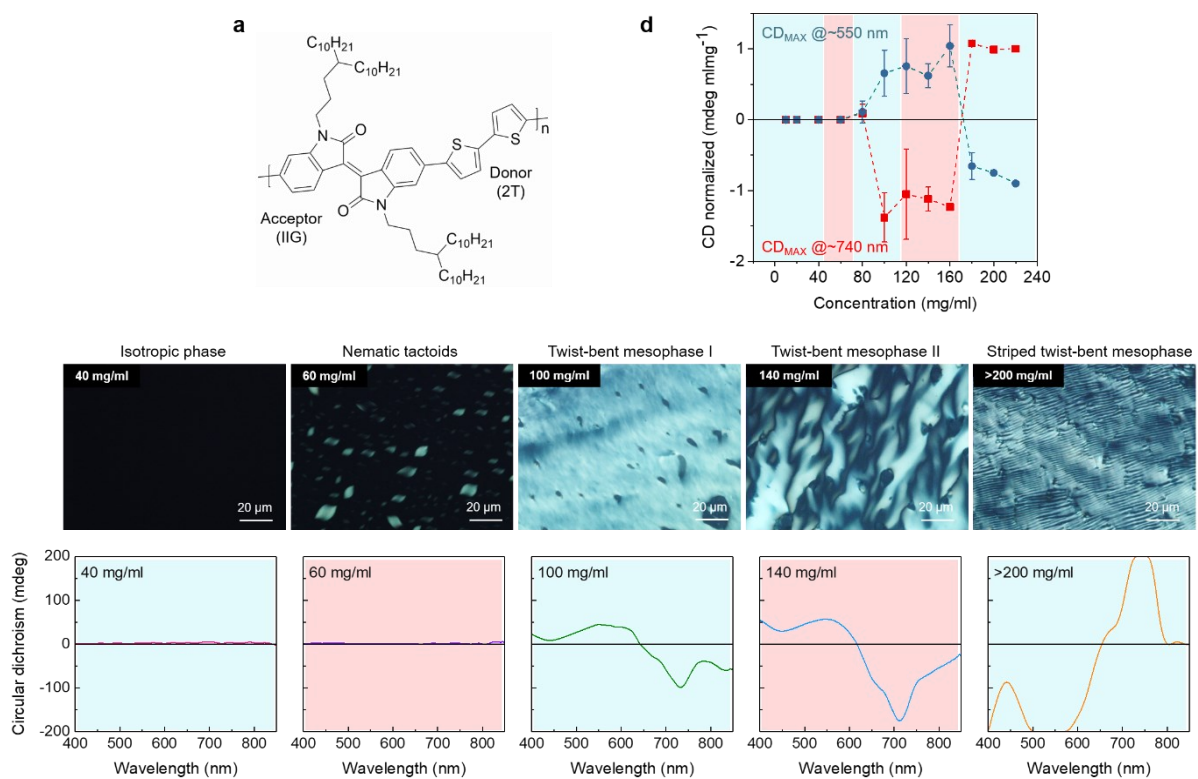

**Figure 1. Chiral emergence of lyotropic LC PII-2T.** (a) Molecular structure of isoindigo-bithiophene–based copolymer (PII-2T). (b) CPOM images of the PII-2T solutions using transmitted light, showing various morphology of the LC mesophases. (c) CD spectra of the corresponding PII-2T solutions shown in c, indicating chirality emerges when exceeding a critical concentration (>60 mg/ml). (d) Averaged maximum values of concentration normalized CD for various solution concentrations where the values are obtained from two main peaks at ~740 nm (red dots) and ~550 nm (blue dots), respectively. The error bars are standard deviations. (c, d) The color code in each panel indicates the five regimes that we observed the distinct morphologies; (i) isotropic phase, (ii) nematic tactoids, (iii) twist-bent mesophase I, (iv) twist-bent mesophase II and (v) striped twist-bent mesophase (a detailed discussion later).

**Mesoscale morphology characterizations.** Further electron microscopy imaging allowed us to explore in detail the nano- and micron-scale structures formed within the mesophase. SEM (Fig. 2a) and TEM images (Fig. 2b) were obtained from samples that were freeze-dried at concentrations corresponding to the entire series of achiral, isotropic to chiral, and twisted mesophases. In the isotropic phase (~40 mg/ml), fibril-like aggregation with a diameter of 40-50 nm was observed. The cross section of fibers was revealed to be nearly circular after measuring the height of individually dispersed fibers (Supplementary Information Fig. S3). In tactoids (~60 mg/ml), the fibers bundle up to form spindle-shaped tactoids with the fiber long axis aligned with curving edge of the tactoid. At ~100 mg/ml, fiber bundles merge into a continuous domain of aligned fibers. The interesting feature in this phase is the locally twisted fibers with a half pitch length of 1.3-1.5 μm. At ~140 mg/ml, the fibers become denser, thicker where the half pitch length becomes shorter (0.7-1.4 μm). As the solution concentration further increases to around 200 mg/ml, a zigzag twinned morphology is observed with a domain width of 0.5-1 μm. This twinned morphology may arise from concerted twisting and bending of densely packed polymer fibers. In addition to micron-scale helicity observed by SEM, high-resolution TEM imaging further unveiled nano-scale helicity of the zigzag twinned morphology (Fig. 2c and 2d). Due to the difficulty of capturing high resolution TEM on relatively thick freeze dried solution samples (~1 μm), printed thin films (~200 nm) that showed resembling morphology were employed instead. Intriguingly, we observed twisted nanoscale fibers (~10 nm) of the twinned morphology, with a helical pitch length of ~170 nm at the domain and with the pitch length of about 80 nm at the boundary. We note here that there should be at least three scales of helicity that span the molecular, nano to micron scales (shown later). We will show through molecular simulations that the nanoscale and micron scale helicity likely arise from the molecular scale helicity.

The electron microscopy imaging clearly reveals that the LC mesophases are comprised of pre-aggregated polymer fibers, rather than single polymer chains. In other words, the mesophases are colloidal, not molecular LCs. Further, the imaging result shows that it is when the fibers become helically twisted that the chirality emerged, at 100 mg/ml. This points to a possible link between chiral emergence and the helical structure of the fiber. With further increasing concentrations, the helically twisted fibers further assemble into higher order structures that take the form of micron-scale twinned domains. Such hierarchical helical assembly of conjugated polymers is reminiscent of biological assembly of helical structures, such as chiral amyloid fibers[1] and chiral M13 phage particles[2]. The helical rod-like shaped M13 phage with a helical pitch length of 3.3 nm assembled into a wavelike "ramen noodle" morphology in a dip-coated thin film. The film exhibited a supramolecular twist pitch length of ~10 μm. Translation of nanoscale helicity to micron scale twists was attributed to both chiral liquid crystalline phase transitions and competing interfacial forces at the meniscus. Also, this periodic morphology has been observed for chiral nematic phase such as cholesteric phase[40, 41], twist-bend nematic phase[42] or heliconical smectic phase[43]. Because there is clearly no molecular chiral center in the polymer and solvent molecules used in this work, the helical mesophases we observed is closer to the twist-bent nematic phase or heliconical smectic phase where the helical assembly originates from symmetry breaking of bent shaped small molecules or colloids. However, as we shall show that the molecular underpinnings of our helical mesophases is distinct from bent-core mesogens.

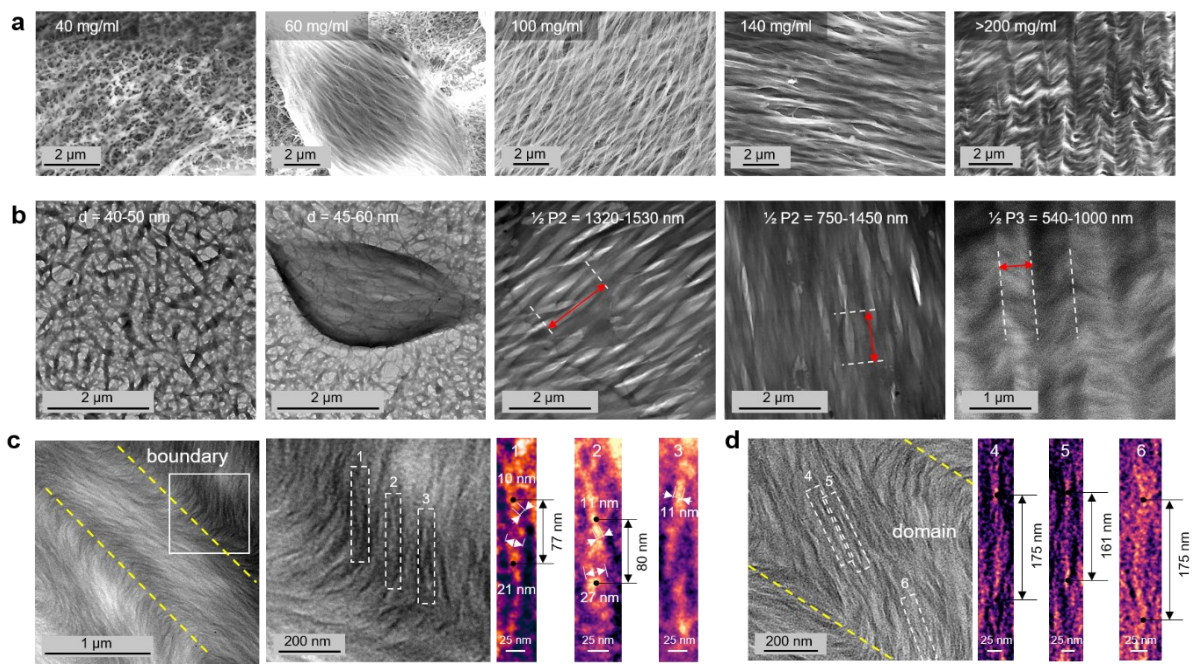

**Figure 2. Micron and nano scale morphology of PII-2T mesophase.** SEM (a, top row) and TEM (b, middle row) images of freeze-dried PII-2T mesophase at increasing solution concentrations (left to right). The half pitch length marked with red arrows in TEM images exhibits a decreasing trend as concentration increases. High-resolution TEM images (c and d, bottom row) of the printed PII-2T film, showing nanoscale twisted fibers on the domain boundaries and within the domains of the twinned thin film. The yellow dash lines denote domain boundaries. 1-6 notes selected regions showing twisted nanoscale fibers. Further imaging analysis was performed using ImageJ.

**Molecular scale structure characterizations.** Next, we determined the internal structure of the polymer fibers constituting the mesophases by characterizing the polymer conformation and intra-fiber molecular packing using small-angle X-ray scattering (SAXS) and grazing incidence wide angle X-ray scattering (GIWAXS). SAXS samples were prepared by the injection of pristine-made solutions into glass capillaries. Figure 3a shows the solution SAXS analysis at various solution concentrations. Note that the SAXS analysis of the solution around 200 mg/ml is absent due to the extremely high viscosity that hinders capillary filling. All solutions show a peak around 0.19-0.22 Å$^{-1}$ which corresponds to lamellar stacking within the polymer nanofibers. This result indicates that the fibers observed from the EM imaging analysis are semicrystalline. Further deconvolution and data fitting provide the lamellar stacking distance inside the fibers and the radius of single polymer chains. The values were obtained through 1D SAXS fitting algorithms reported in our recent work[44]. Briefly, both pre-aggregates and dispersed polymer chains co-exist in chlorobenzene solutions with the lamellar peak resulting from lamellar stacking of polymer chains within the pre-aggregate. Therefore, our model consists of a power law, semiflexible cylinder, and pseudo-Voigt peak to fit the contributions from the pre-aggregates, polymer chains, and lamellar peak, respectively. In all samples, a low q power law exponent of around –3.4 was observed, indicating Porod scattering from large pre-aggregates which is in agreement with our observation that aggregated polymer fibers constitute the mesophases. The semiflexible cylinder and pseudo-Voigt peak were used to extract the radius of the polymer cross-section and the lamellar stacking distance, respectively. Figure 3b shows the lamellar stacking distance increases as solution concentration increases. To explain this, we propose that as polymer chains become more twisted, the effective volume of polymer chains may be increased and thus molecular packing in the fibers may be loosened, leading to increased lamella stacking distance. The radius of the polymer chain cross section (Fig. 3c) closely matches with half lamellar distance in most cases, indicating absence of side chain interdigitation in all the solution phases. We note that typical lamellar stacking distance of solid state films is ~25 Å, which is smaller than the values (28-33 Å) we obtained from solution phases. We attribute larger lamella stacking distance in solution phases to more disorder and/or more expanded/swollen side chain conformation due to the presence of solvent.

We then utilized GIWAXS to characterize molecular stacking and orientation distribution within these mesophases. Freeze-dried solution samples used for imaging analysis were investigated through GIWAXS. Two-dimensional (2D) GIWAXS patterns of the mesophases are provided in Fig. 3d. We observed the (010) π-π stacking peaks for all samples, which affirms the semi-crystallinity of polymer fibers. Molecular packing details for the π-π and lamellar stacking peaks are summarized in Supplementary Information, Table S1. With increasing solution concentration, the π-π stacking distance progressively reduced from a range of 3.61-3.65 Å for the isotropic phase to 3.56-3.59 Å for the twist-bent mesophase II (Fig. 3e). This indicates that increased backbone torsion has afforded closer π-π stacking, or stronger polymer interactions in highly concentrated solutions have led to more torsional backbone. This connection between backbone torsion and π-π stacking has been reported in earlier works[45, 46]. Pole figure analysis[47] (Supplementary Information Fig. S4) performed on the π-π and lamellar stacking peaks as a function of the polar angle ($\chi$) provides indirect evidence of twisted nanofibers in the mesophases through molecular orientation distributions. A distinct preferential edge-on orientation of π-stacks is observed in the isotropic phase (10 mg/ml) with a minor contribution from face-on π-stacks (Fig. 3f, left). The degree of out-of-plane orientation was quantified using the 2D orientation parameter, $S_{2D} = 2\langle\cos^2 \gamma\rangle - 1$, where $\gamma$ is the tilt angle of the conjugated backbone with respect to the substrate (Fig. 3f, middle). Values of $S_{2D} = 1, 0$ and $-1$ correspond to perfect edge-on, isotropic and perfect face-on orientation, respectively. With the appearance of mesophases at and beyond ~60 mg/ml, the orientation of π-stacks becomes more isotropic or weakly face-on, which we believe is associated with twisted polymer backbone in helical mesophases.

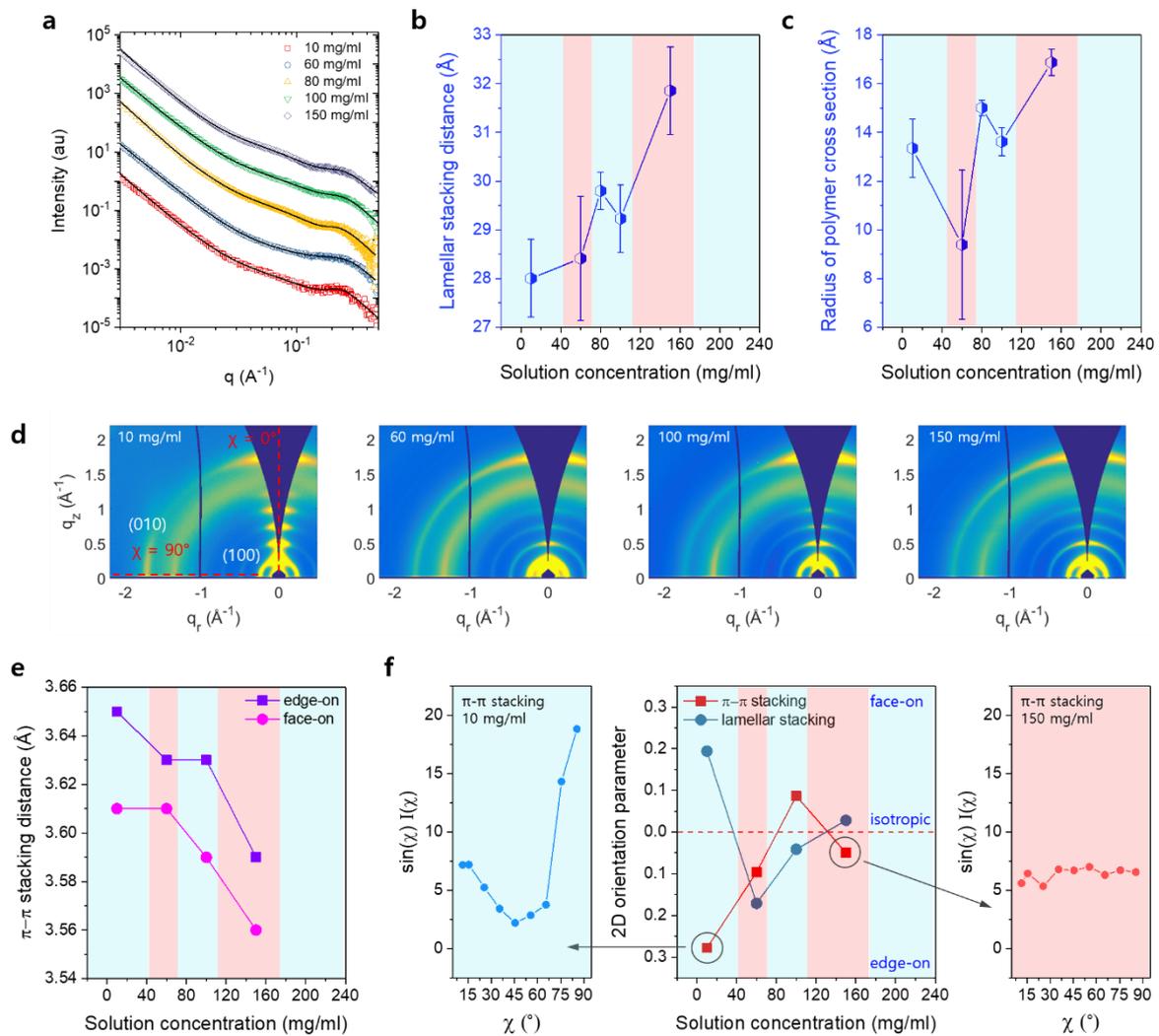

**Figure 3. Molecular scale assembly of PII-2T mesophase.** (a) SAXS plots of PII-2T mesophase with increasing solution concentration (as labeled). Black solid lines are fitting results using the model developed in ref 44. (b) Lamellar stacking distance obtained from the fitting, showing an increasing trend as solution concentration increases. (c) Radius of the cross section of a single polymer chain extracted from the Guinier knee near 0.07 Å$^{-1}$. (d) 2D X-ray scattering patterns for the freeze-dried mesophases prepared at the indicated concentration. (e) Face-on and edge-on π-π stacking distance of the mesophases at various solution concentrations. (f) Out-of-plane 2D orientation parameter of the mesophases (middle). $S_{2D}$ value closer to -1 or 1 indicates an edge-on and face-on orientation, respectively and $S_{2D} = 0$ indicates an isotropic orientation. Geometrically corrected intensity of π-π stacking (010) peak as a function of polar angle, χ for isotropic phase at 10 mg/ml (left) and twisted mesophase at 150 mg/ml (right). Note that χ = 0° and 90° correspond to face-on and edge-on orientation, respectively. (b, c, e, f) The blue and red color codes in each panel indicate the five distinct solution phases outlined in Fig. 1.

We further inferred molecular conformations of these mesophases by UV-Vis absorption spectroscopy. According to our linear POM and polarized UV-Vis spectroscopy (Supplementary, Fig. S5-S6), the polymer chains are aligned along the fiber long axis with the optical transition dipole moment occurring mainly along the polymer backbones. Figure 4a shows UV-Vis absorption spectra of PII-2T solutions that correspond to the series of isotropic to twisted mesophases. There are several main peaks positioned around 715 nm, 650 nm, 420 nm and broaden peaks around 500-600 nm. The lowest energy peak around 715 nm is mainly attributed to the intramolecular π-π* transition from the pre-aggregates. The second lowest energy peak around 650 nm results from the intermolecular transition in the pre-aggregates. These two peaks markedly decreased as the aggregates dissolved into the solution at increased temperature, confirming that the two peaks are associated with pre-aggregates temperature (Supplementary Information Fig. S7). The far higher energy transition around 400 nm is attributed to localized transition from isoindigo units[33] which becomes dominant when the polymer chains are more torsional and the electrons are more localized. In other words, the optical transition at 715 nm and 650 nm can be considered as contributions from J- and H-aggregation inside the pre-aggregated polymer fibers, respectively. Thus, the increase in absorbance ratio $A_{650}/A_{715}$ as well as the sharp decline of absorption coefficient with increasing solution concentration suggest a decrease in π-conjugation and increase in backbone torsion from isotropic to twisted mesophase (Fig. 4b). Moreover, the two main peaks (715 nm and 650 nm) are blue-shifted with increasing solution concentration (Fig. 4c), showing that the transition energy increases owing to decreasing conjugation length caused by torsional polymer chains.

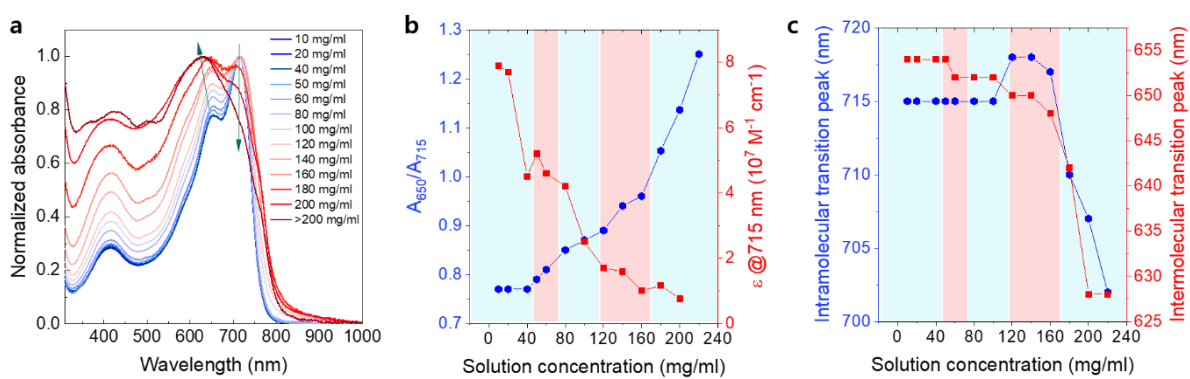

**Figure 4. Optical transition property of PII-2T mesophase.** (a) UV-Vis absorption spectra of PII-2T solutions that corresponds an entire series of achiral, isotropic to chiral, twisted phases. The arrows indicate peak changes with change increasing solution concentration. (b) The two peak ratio ($A_{650}/A_{715}$) and absorption coefficient ($\varepsilon$) change indicating twisted molecular conformation as concentration increases. (c) Intramolecular and intermolecular transition peak position changes, showing that the aggregates are more twisted and/or polymer conformation is more torsional as well. The blue and red color codes in b and c indicates the five distinct phases corresponding the morphology transition observed above.

**Molecular simulations unveil helical conformation.** We next performed MD simulations on PII-2T to answer how structural features at the molecular level lead to chiral emergence. We explore two possible molecular origins: first, polymer chains may adopt wavy, helical conformation arisen from asymmetric dihedral potentials; second, molecular stacking may occur in a staggered fashion leading to chiral assembly. The latter point was shown in three other systems, where chiral aromatic peptide[48], chiral nanoplatelets[49] and achiral nanocubes[8] were found to stack at an angle giving rise to chiral helical structures.

To determine polymer conformation in solution, we constructed a system containing a 30-mer of PII-2T surrounded by chloroform and simulated it without any bias for ~260 nanoseconds. The conformation of the entire polymer chain fluctuates with multiple regions exhibiting local helical structure (Fig. 5a; Supplementary Information, Movie 2). The half pitch of the helix is comprised of 5 to 6 monomers corresponding to a length of about 60 – 90 Å. To understand the molecular underpinnings of this intriguing phenomenon, we closely examined the distribution of four dihedral angles at various positions along the 30-mer (Supplementary Information, Fig. S8 and S9). The four dihedral angles are illustrated in Figure 5b: the angle between the thiophene-isoindigo (T-I), the internal angle in the isoindigo unit (I-I), the angle between the isoindigo-thiophene (I-T) and the angle between the thiophene-thiophene (T-T). The I-I dihedral angle is the most rigid, with the narrowest distribution centered around 25° (Supplementary Information, Fig. S8). The I-T and T-I dihedrals fluctuate between four well-defined positions (two cis and two trans conformations), all deviating from co-planarity by ~30°. In contrast, the T-T dihedral angle is the most flexible and fluctuates between +150 and –150°, with two peaks around +/–90 degrees (Fig. 5c). This suggests that the T-T dihedrals may act as flexible hinges, enabling the formation of wavy polymer conformation. Indeed, we observed a small but discernable preference towards cis conformation of T-T dihedrals in curved regions along the polymer chains, where two successive thiophene rings are oriented towards the same direction (Fig. 5d, Supplementary Information, Fig. S10). Moreover, we noticed that the distribution of T-T dihedral angle is unbalanced on each side of the zero degree (Fig. 5c). This indicates an asymmetry in the dihedral potential, which may serve as a source for symmetry breaking during chiral assembly.

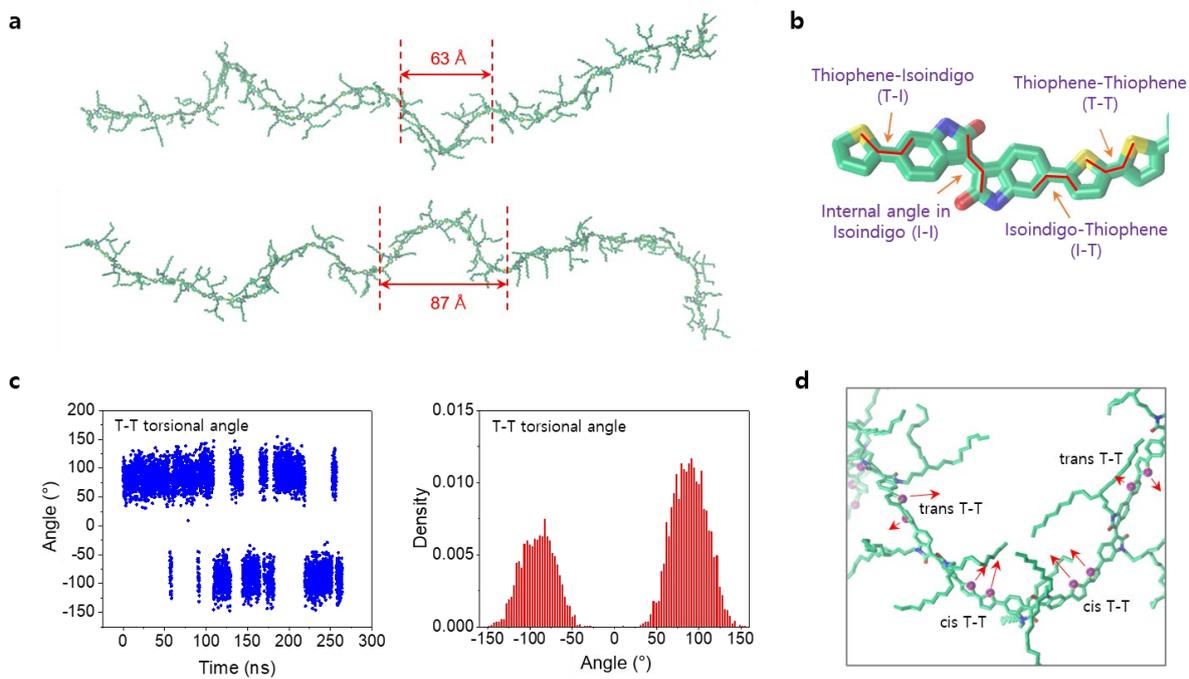

**Figure 5. Helical conformation of PII-2T molecules in a solution phase.** (a) MD simulation of a 30-mer of PII-2T in chloroform, showing helix-like flexible structures with a half pitch length of about 60–90 Å. (b) PII-2T molecular structure (omitted alkyl side chains and hydrogens) with each bond used in dihedral angle frequency plots. (c) T-T torsion angle plot between the 14th and 15th monomers of the 30-mer (left) and the corresponding angle distribution histogram (right). (d) Enlarged structure of the curved region on the helix structure. The purple dots are sulfur atoms on the thiophene rings. The arrow marks indicate a local conformation of the thiophen rings, showing a preferential cis-conformation in the curved region.

To understand whether asymmetric molecular stacking contributes to structural chirality, we simulated dimeric assembly of PII-2T oligomers in chloroform. To reduce the computational cost, we substituted the long alkyl side-chains ($C_{24}H_{49}$) with short ones ($C_4H_9$) and constructed hexamers to better observe the interaction between backbones. During the ~400 ns simulation (Supplementary Information Movie 3), the hexamers that were 20 Å apart at the beginning took ~100 ns to form a dimer with their backbones aligned in a stable parallel conformation (Fig. 6a). The interaction between polymer backbones comes from the stacking between their thiophene and indole rings. To further explore the thermodynamics of interaction between backbone aromatic rings, we performed umbrella sampling simulations[50] to obtain the potential mean force (PMF) curve for dimerization between the aromatic rings. The center-of-mass distance (I) and the dihedral angle (θ) between the facing indole ring and the bithiophene rings of the two hexamers were used as two reaction coordinates (Fig. 6b). First, Umbrella sampling was performed varying I from 3 to 13 Å using a window size of 0.5 Å. The PMF (Fig. 6c) showed that the free energy of dimerization reached the minimum at ~4.6 Å, which closely matches with the average π-stacking distance (~4.4 Å) in amorphous regions of the thin films measured by GIWAXS[33]. We also performed Umbrella sampling along θ and the PMF curve showed the free energy minimum at ~45° (Fig. 6d), which is comparable to the dihedral angle observed in the unbiased simulation. Intriguingly, the potential well around the ~45° minimum was found to be asymmetric. Such staggered, asymmetric dimeric stacking likely plays a role in inducing structural chirality as well, besides helical conformation arisen from asymmetric torsional potential. Indeed, we spotted in our simulation consecutive twist and bent oligomers that are dimerized in a helical fashion via the indole-thiophene stacking with a half pitch length of ~50 Å (Fig. 6e).

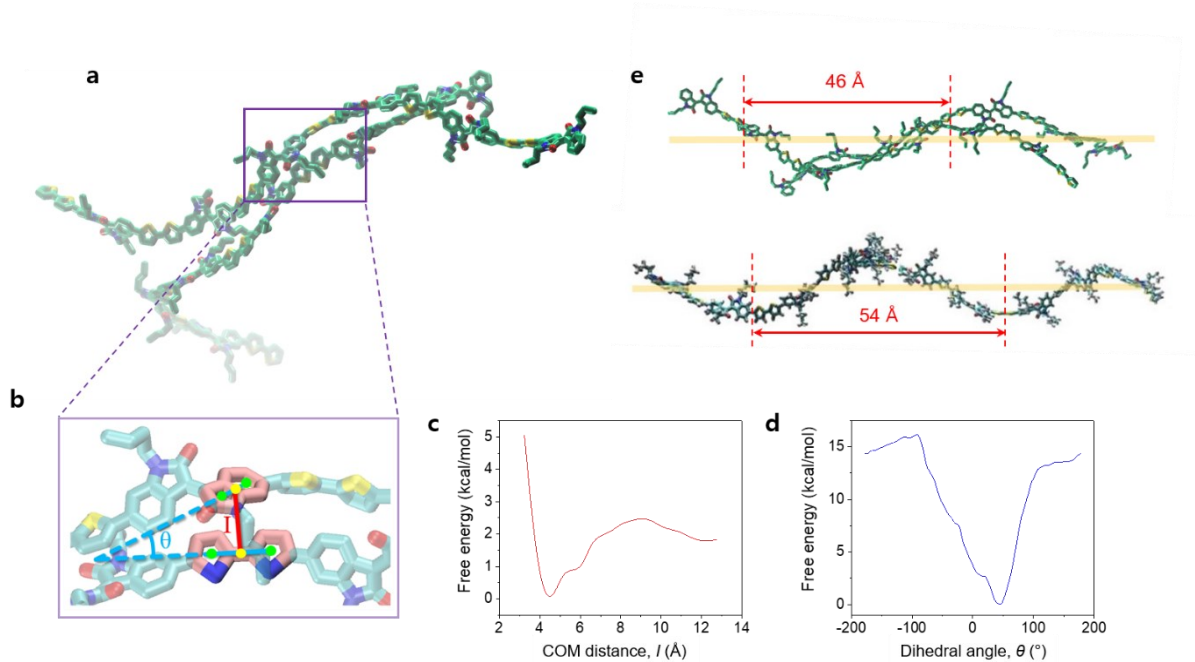

**Figure 6. Helical structure of stacked PII-2T molecules in the solution phase.** (a) Dimeric assembly of PII-2T hexamers captured from the simulation, showing the aromatic rings stacked closely. (b) Reaction coordinates used for umbrella sampling. Red stick indicates the center of mass distance, l between the bithiophene and indole rings (the center of mass denoted by yellow dots). Blue stick and dashed line indicate the dihedral angle, θ formed by the center of mass of these four rings (the center of mass denoted by green dots). (c) PMF plot as a function of the distance, l obtained from umbrella sampling simulation using Weighted Histogram Analysis Method (WHAM), showing the minimum free energy at ~4.6 Å. (d) PMF plot as a function of the angle, θ, showing the minimum free energy at ~45°. (e) Stacked oligomers captured from the simulation, showing the helical molecular assembly with a pitch length of about 50 Å.

**Hierarchical morphology of chiral mesophases.** All above characterizations culminate in a multiscale morphology model to reveal the chiral emergence of achiral conjugated polymers via a multistep hierarchical assembly process (Fig. 7). In dilute solutions, single polymer chains are prone to adopting wavy helical conformation owing to flexible dihedrals that easily accommodate curvature modulation via cis-trans transition. Despite globally achiral, local chiral conformation may exist transiently at the single molecule level arisen from asymmetric torsional potential of the flexible dihedrals. Upon aggregation, polymer backbones π-stack in a staggered fashion forming helical nanofibers of 40-50nm in diameter with a pitch length of 10-20nm. Due to the asymmetry of the free energy well, such π-π stacking may give rise to local structural chirality at the single fiber level, while the ensemble average over all fibers remains achiral. The nanofibers constitute the isotropic solution until nematic tactoids nucleate at ~50mg/ml wherein nanofibers aggregate into spindles confined in mesophase droplets. The absence of global chirality may be explained by the symmetric shape of the spindles or the statistical distributions of chirality from droplet to droplet. Increasing concentration to 100mg/ml eventually causes nematic tactoids to merge into a single coherent mesophase – twist-bent mesophase I, while at the same time breaking chiral symmetry. Helical, twisted fiber bundles constitute this mesophase, forming uniformly aligned and birefringent domains over large area; this suggests long range interactions between fibers to induce single handedness arrangement. Further increasing concentration densifies the fibers, thickens the fiber bundles and reduces the helical pitch in twist-bent mesophase II. Ultimately, striped twist-bent mesophase appears where densely packed fibers twist and bend coherently to result in micron scale zigzag twinned domains. Overall, increasing concentration enhances backbone torsion at the molecular scale, brings closer π-π stacking while loosens lamella stacking at the mesoscale, and decreases helical pitch length at the microscale according to UV-vis, GIWAXS, SAXS, and EM analysis. The structural changes at the molecular scale probably underlie the structural transitions observed. In summary, our analyses suggest that chirality emerged from asymmetrically torsional polymer conformation, asymmetrically staggered backbone stacking, and their coherence over long range when exceeding a critical concentration.

Is the observed emergence of chiral mesophases unique to PII-2T or more general among conjugated polymers? Interestingly, similar zigzag twinned morphology has been previously observed for several well-studied conjugated polymer systems, for instance, poly(3-hexylthiophene) (P3HT)[51], poly-2,5-bis(3-alkylthiophen-2-yl)thieno[3,2b]thiophenes

(pBTTT)[52], naphthalene diimide-bithiophene-based copolymer (P(NDI2OD-T2))[27, 32] and diketopyrrolopyrrole-benzothiadiazole-based copolymer (DPP-BTz)[33, 53]. These polymers were also found to have a lyotropic LC nature. Zone-cast pBTTT has exhibited aligned micron scale domains with alternating backbone orientation tilted by about 45° [52]. Meniscus-guided coated P(NDI2OD-T2) and DPP-BTz films also showed the zigzag twinned morphology in which the wave-like micron scale structures formed along the coating direction[32, 33, 53]. Aged P3HT solution confined in a rectangular capillary has exhibited liquid crystalline orders with similar alternating dark and bright stripes that oriented perpendicular to the capillary long axis[51]. While reported before, such zig-zag morphology of conjugated polymer thin films has not been previously associated with chiral mesophases in previous reports, nor are the multiscale structure explored. On the other hand, this unique morphology seems to be strongly associated with structural chirality as similar zig-zag twinned morphology was formed by chiral colloidal particles (M13 phage)[2] or bent-core mesogens[54]. Previous reports on similar film morphologies from multiple classic conjugated polymer systems suggest that our observed chiral emergence in lyotropic liquid crystal phases may be more general.

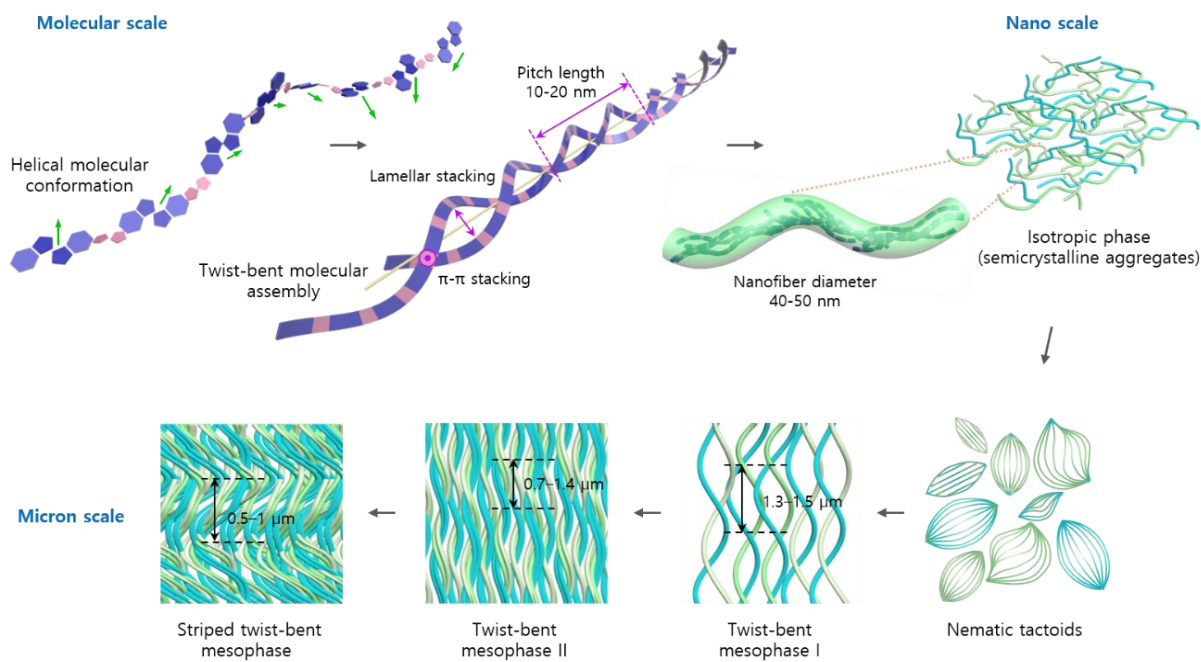

**Figure 7. Schematic illustration of proposed chiral emergence in multistep hierarchical assembly of achiral conjugated polymers.**

**Conclusion**

In summary, we report helical structures of achiral conjugated polymers that were developed through the multistep hierarchical assembly pathways for the first time. Through a combination of in-situ and ex-situ optical and electron microscopy, along with X-ray scattering, we observed that LC mesophases evolve from pre-aggregated polymer nanofibers to assemble into chiral LC mesophases as the fibers become more twisted and bundle up in a helical fashion. Our combined experimental and computational analyses suggest that chirality emerged from asymmetrically torsional polymer conformation, asymmetrically staggered backbone stacking, and their coherence over long range when exceeding a critical concentration. Thus, by considering the role of achiral-to-chiral transition through hierarchical assembly, we are able to fully explain the structural evolution of this conjugated polymer across concentration regimes spanning the dilute to thin-film states – an assembly pathway underlying all evaporative solution coating and printing processes. This study provides a clear multiscale picture of the structural origin of chiral emergence – a phenomenon general among multiple classes of conjugated polymers which has been overlooked before. We anticipate that the ability to precisely control chiral structures of semiconducting and conducting polymers will open avenues to exciting new optical, electronic, spintronic, mechanical, and biological properties not possible before.

**Acknowledgement**

K.S.P. and Y.D. acknowledge ONR support under Grant No. N00014-19-1-2146, and K.S.P. acknowledges partial support from the Shen Postdoctoral Fellowship. B.B.P. and Y.D. acknowledge NSF DMREF for funding, under Grant Number 17-27605. J.J.K. and Y.D. acknowledge support by the NSF CAREER award under Grant No. 18-47828. J.J.K. also acknowledges partial support from the U.S. Department of Energy, Office of Science, Office of Workforce Development for Teachers and Scientists, Office of Science Graduate Student Research (SCGSR) program. The SCGSR program is administered by the Oak Ridge Institute for Science and Education for the DOE under Contract Number DE-SC0014664. P.K. acknowledges partial support by the NSF MRSEC: Illinois Materials Research Center under Grant Number DMR 17-20633, the 3M Corporate Fellowship, and the Harry G. Drickamer Graduate Research Fellowship. H.A. and Q.C. acknowledge support from Air Force Office of Scientific Research grant AFOSR FA9550-20-1-0257.

**Author Contributions**

K.S.P. and Y.D. designed the research project, and Y.D. supervised the project. K.S.P. carried out the experiments and analyzed the data. Z.X. performed the MD simulations and drafted the MD writing under the supervision of D.S. B.B.P developed the freeze-drying set-up with K.S.P. and performed the SEM measurement. H.A performed the TEM measurements and analyzed the data under the supervision of Q.C. J.J.K. performed the SAXS measurements and data fitting and drafted the SAXS writing. P.K. performed the GIXD measurements. K.S.P. and Y.D. wrote the manuscript. All authors discussed, revised, and approved the manuscript.

**Competing Interests statement**

The authors declare no competing interests.

**Materials and Methods**

**Materials**

The isoindigo-based copolymer, PII-2T [number-average MW (Mn) = 30,645 g/mol, weight-average MW (Mw) = 76,809 g/mol, and polydispersity index (PDI) = 2.50] was synthesized as previously described[55]. The solution was prepared by dissolving the polymer (10 to 150 mg/ml) in chlorobenzene (CB; anhydrous, 99.8%; Sigma-Aldrich Inc.). A bare-Si and micro cover glass were used as a bottom substrate and top cover, respectively. Corning glass substrates were used for spectroscopic studies. The substrates were cleaned with toluene, acetone, and isopropyl alcohol and then blow-dried with a stream of nitrogen to remove contaminants. Poly(sodium 4-styrenesulfonate) (PSS) average Mw ~70,000, powder was used as a sacrificial layer to transfer the polymer films for TEM characterization.

**Solution-state sample preparation and characterization**

All solution samples except the ones for SAXS were prepared by a drop-and-dry method. The drop-and-dry method is basically concentrating the pristine solution by adding multiple numbers of solution droplets, drying them out and blending with a solution drop. Briefly, a needed concentration was obtained from casting and drying a multiple number of the solution drops on a local spot of the substrate first and blending/shearing with the last drop of the stock solution (2 μl). The sandwiched solution samples were further run through thermal annealing cycles to reach an equilibrium state. We observed the crystalline mesophase beyond a critical concentration both under nonequilibrium droplet-drying conditions and at near-equilibrium conditions after thermal treatments. Selected high concentrated pristine-made solutions (60–150 mg/ml) showed resembling morphologies of crystalline mesophases compared to samples prepared by the droplet-drying method (Supplementary Fig. S11). This confirms that our droplet-drying method is reliable to study concentration dependent solution phases even without demanding a large amount of materials. For instance, a 50 mg/ml solution was made with dried four drops of 2 μL 10 mg/ml solution on a bare Si substrate and subsequently blending by additional one drop of 2 μL 10 mg/ml solution with a glass coverslip. In order to reach an equilibrium state, the sandwiched sample was run through a moderate heating and cooling process (25°C→100°C→25°C). The birefringence of mesophases was observed using CPOM (Eclipse Ci-POL, Nikon). The samples for spectroscopy were prepared with the same method using corning glass substrates and glass coverslips. UV-vis (Cary 60 UV-Vis, Agilent) spectroscopy was used to calculate the absorption coefficient and investigate polymer conformation in the solution phase. CD spectra were recorded using a JASCO J-810 spectrophotometer. To eliminate contributions from linear dichroism and birefringence, all samples were investigated at numerous sample rotation angles. Identical spectra were recorded at different sample batches demonstrating the chiral nature of the phase.

**Solid-state sample preparation and characterization**

A freeze-drying method was performed to characterize the solution state through electron microscopy tools. The structure of aggregates in solution was imaged using scanning electron microscopy (SEM) and transmission electron microscopy (TEM). The samples sandwiched between a bare Si substrate and a glass coverslip was first submerged in liquid nitrogen. The top coverslip was then removed inside the liquid nitrogen bath. The sample was immediately transferred to a sealed Linkam thermal stage chamber (LTS420) which is held at -100 °C in a nitrogen atmosphere. The temperature was slowly (0.5 °C/min) raised to -80 °C for chlorobenzene sublimation under the vacuum. It took ~6 hours to fully sublimate the solvent. Finally, the temperature was raised to 25 °C. The entire procedure was able to be monitored under the microscope, which ensured us to carefully preserve the solution states. The prepared samples were imaged using SEM (JEOL JSM-7000F at 25 kV accelerating voltage) and TEM (JEOL 2100 Cryo TEM with a $LaB_6$ emitter at 200 kV). For TEM imaging, low electron dose rates (4–12 $e^-$ $Å^{-2}$ $s^{-1}$) were applied using spot size 3 to minimize beam-induced alteration. Each image was collected with an exposure time of 1 s, resulting in a dose per image of 4–12 $e^-Å^{-2}$. A defocus of $-10,240$ nm was used throughout all image acquisition to improve contrast. For TEM characterization, all procedure was performed same using PSS layer deposited on the Si substrate. 10 wt% PSS in water solution was spin coated on the Si substrate at 5000 rpm for 1 min. The freeze drying polymer films on the PSS was transferred on copper grids (Ted pella, 01840-F) in a water bath. Printed PII-2T films were prepared onto PSS-coated substrates by a blade coating method. Briefly, an OTS-treated Si substrate was used as a blade set at an angle of 7°, with a gap of 100 μm between the substrate and the blade. The blade was linearly translated over the stationary substrate while retaining the ink solution within the gap. The PII-2T films were printed on PSS-coated Si substrates at printing speeds at 0.2 mm/s with a substrate temperature of 65 °C. The polymer solutions was 5 mg/ml dissolved in chlorobenzene. The printed films on the PSS was transferred on copper grids in a water bath. The freeze-drying samples prepared on $SiO_2$ were also measured using GIWAXS. GIWAXS measurements were performed at beamline 8-ID-E at the Argonne National Laboratory, with an incident beam energy of 7.35 keV on a 2D detector (PILATUS 1M) at a 208-mm sample-to-detector distance. Samples were scanned for 10 s in a helium chamber. The x-ray incident angle was set to be above (0.14°) the critical angle (≈0.1°) of the polymer layer (penetration depth, approximately 5 nm). We note that the GIWAXS analysis of the solution around 200 mg/ml is lacking because of the small sample area (a few hundred micrometers) when compared to the beam irradiated area (~5 mm).

**SAXS experiments and analysis**

SAXS experiments were carried out at the 12-ID-B beamline of the Advanced Photon Source at Argonne National Laboratory using an X-ray beam energy of 13.3 keV. A Pilatus 2M detector was used primarily at a sample-to-detector distance of 3.6 m. The polymer solution SAXS experiments were performed using a flow cell to prevent beam damage and enable longer exposure times. The flow cell was constructed using a 1 mm diameter quartz capillary connected to PTFE tubing using PTFE heat shrink tubing. The tubing was connected to a syringe pump which cycled the polymer solution at a linear velocity of about 1 mm/s within the capillary while a series of 0.1 s exposures with 3 s delays were accumulated. The isotropic 2D scattering patterns were averaged, reduced, and then background subtracted using the beamline's MATLAB package. The 1D scattering profiles were then analyzed and fit using custom models in SasView.

**Molecular Dynamics Simulation**

<u>System setup.</u> Topology files of polymers were constructed using Antechamber[56]. The initial coordinate files of the systems were generated using Packmol[57]. The general AMBER force field (GAFF)[58] was used for parameterizing the polymer. The coordinate files of polymers were obtained by drawing and saving as .pdb using PubChem Sketcher[59]. The polymer molecules were solvated in boxes of TIP3P[60] chloroform molecules.

Simulation and data analysis. All simulations were run in Amber18[61] with a time step of 2 fs and with hydrogen-containing bonds constrained using the SHAKE algorithm[62]. All production runs were maintained at a constant temperature of 300 K using a Langevin thermostat with a coupling time constant of 2 ps and at a constant pressure of 1 bat using a Berendsen barostat with a τ of 2 ps. A cutoff of 10 Å was used for nonbonded interactions. Frames were saved to trajectory files every 1 ns.

All umbrella sampling simulations were run in Amber18[61]. Amber's harmonic restraints were used for restricting the distance or dihedral angle. In the umbrella sampling using COM distance as a collective variable, we used umbrella windows spaced 0.5 Å apart, and a force constant of 20 kcal/mol Å. The umbrella windows were covering the distance from 1 to 13 Å, and the simulation was run on each window for 10 ns. During the following data analysis, since the distance could not go below 3 Å, that part of the data was cut. In umbrella sampling using dihedral angle as a collective variable, we used umbrella windows spaced 3 degrees apart, and a force constant of 200 kcal/mol rad$^2$. The umbrella windows were covering dihedral angles from -180 to 180 degrees, and the simulation was run on each window for 2 ns.

Weighted Histogram Analysis Method (WHAM) was used to analyze the data from umbrella sampling simulation and generate the PMF[63, 64]. The PMF was generated with an assumed temperature at 300 K and no padding. And the statistical error was estimated using Monte Carlo bootstrap error analysis, with 10 fake data sets generated using a random seed of 5. For more details, please see Supplementary Methods.

# Supplementary Information

Kyung Sun Park[1], Zhengyuan Xue[1], Bijal Patel[1], Hyosung An[2], Justin J. Kwok[2], Prapti Pkafle[1], Qian Chen[2], Diwakar Shukla[1] and Ying Diao[1,2,3]*

[1]Department of Chemical and Biomolecular Engineering, University of Illinois, Urbana, USA

[2]Department of Materials Science and Engineering, University of Illinois, Urbana, USA

[3]Beckman Institute, Molecular Science and Engineering, University of Illinois, Urbana, USA

*Corresponding author. Email: yingdiao@illinois.edu

**MD simulation details**

MD simulation on long PII-2T chain (30mer) was performed to investigate the conformation of actual polymer chains in chloroform solvent. Moreover, simulations on PII-2T oligomers (hexamer) with removal of alkyl chains were performed to investigate the behavior of PII-2T oligomer in chloroform and explore the source of assembly and chirality emergence. Lastly, umbrella sampling simulations were performed to study the relationship of interaction between backbones and position of the aromatic rings. Simulation details are summarized in the table below:

| System | Polymer length | Polymer position | $CHCl_3$ | Simulation Box (Å) | Simulation Time (ns) |
|---|---|---|---|---|---|
| 30-unit PII-2T chain | 30 | Fixed | 29233 | 104.0*99.0*448.6 | 262 |
| 2 hexamer with removal of alkyl chains | 6 | Dynamic | 4534 | 90.9*114.5*75.0 | 400 |

**Initial Setup of 30-unit polymer simulations.** The starting structures of PII2T monomers were drawn using PubChem Sketcher[1] and optimized by the built-in structure optimization function of Avogadro[2]. The monomer topology file was generated by Antechamber and using the general AMBER force field (GAFF)[3]. The 30mer chain structure model was generated by repeating the monomer structure for 18 times in Antechamber. Periodic simulation box of 448.6 Å in length with the central ax alongside with the 30mer was generated using tLEaP from AmberTools. The system was then solvated with 29233 $CHCl_3$ molecules, and topology file was generated using tLEaP[3].

**Initial Setup of cut side-chain hexamer simulations.** The starting structures of PII2T monomer with removal of alkyl chains was generated using PyMol[4]. The hexamer chain structure model was generated by repeating the monomer structure for 6 times in Antechamber[3].

System with periodic simulation box of the size 90.9*114.5*75.0 Å and 2 hexamer putting parallelly with 20 Å apart at the center was generated using Packmol[5]. The system was then solvated with 4534 $CHCl_3$ molecules in tLEaP[3].

All the simulations were set up using the AMBERTools18 and performed with AMBER18 software using the general AMBER force filed (GAFF)[3]. All the partial charges were derived using the AM1/BCC method. The same partial charges from monomer PII-2T were used for each unit of the PII-2T polymer chain. Previous studies have shown that the GAFF can accurately reveal behaviors of PII-2T in chloroform solvent[6]. Both the systems was minimized using steepest descent method for 9200 steps and then slowly heated up to 300 K in 20 ps before simulation.

All the simulations were performed in NPT ensemble (1 atm, 300K) with periodic boundary conditions. Particle-mesh Ewald method was used to treat the electrostatic interactions with a 10Å cutoff distance[7]. The SHAKE algorithm was applied to constrain the length of covalent bonds involved hydrogen atoms to their equilibrium values[8]. The integration step was 2 fs. Berendsen thermos-barostat with a damping time constant of 2 ps was used to control the temperature and pressure of the ensembles[9].

**Umbrella Sampling Simulation and Data Analysis.** All umbrella sampling simulations were run in Amber18[3], with same basic setting as previous simulations. Amber's harmonic restraints were used for restricting the distance or dihedral angle. In the umbrella sampling using COM distance as a collective variable, we used umbrella windows spaced 0.5 Å apart, and a force constant of 20 kcals/mol Å. The umbrella windows were covering the distance from 1 to 13 Å, and the simulation was run on each window for 10 ns. During the following data analysis, since the distance could not go below 3 Å, that part of the data was cut. In umbrella sampling using dihedral angle as a collective variable, we used umbrella windows spaced 3 degrees apart, and a force constant of 200 kcal/mol rad$^2$. The umbrella windows were covering dihedral angles from -180 to 180 degrees, and the simulation was run on each window for 2 ns.

Weighted Histogram Analysis Method (WHAM) were used to analyze the data from umbrella sampling simulation and generate the PMF[10, 11]. The PMF were generated with assumed temperature at 300 K and no padding. And statistical error was estimated using Monte Carlo bootstrap error analysis, with 10 fake data sets generated using random seed of 5.

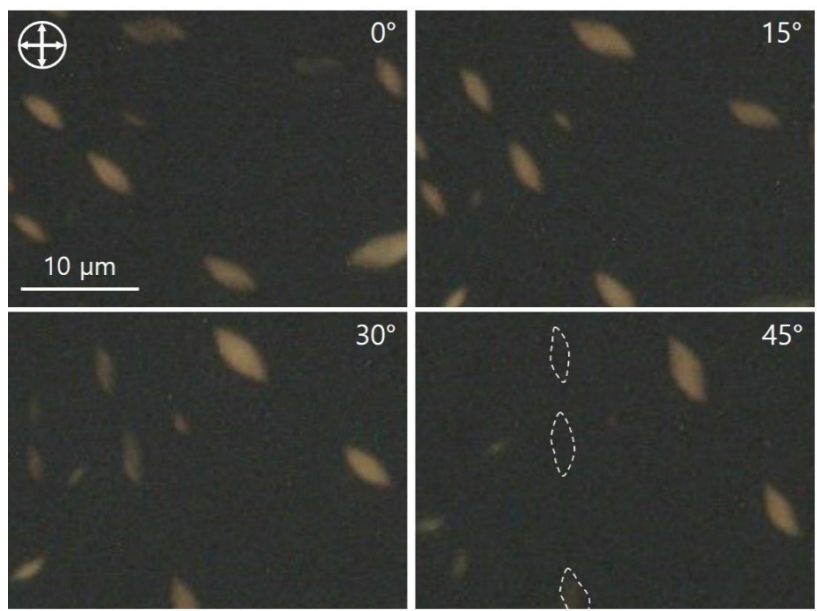

**Figure S1. Homogeneous tactoids observed at 50 mg/ml PII-2T solution.** The major axis rotated clockwise with respect to a polarizer by 0◦, 15◦, 30◦, and 45◦. Homogenous tactoids show a uniform change in brightness when the tactoid is rotated from 0◦ to 45◦ relative to either of the polarizers, and are uniformly dark at 45◦.

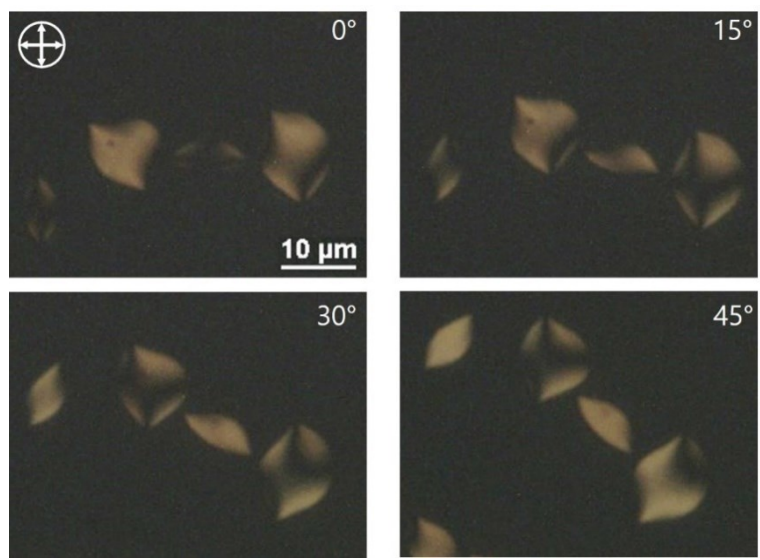

**Figure S2. Bipolar tactoids observed at 60 mg/ml PII-2T solution.** The major axis rotated clockwise with respect to a polarizer by 0◦, 15◦, 30◦, and 45◦. Bipolar tactoids show four dark brushes crossing at the center when either of the crossed polarizers is aligned with the major axis of the tactoid. By rotating the tactoid, the dark cross splits into two dark curved brushes that move away from each other.

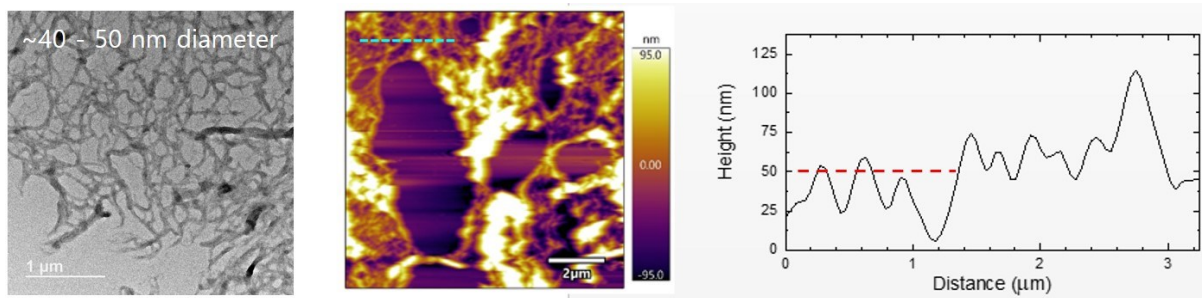

**Figure S3. TEM and AFM images of 10 mg/ml PII-2T freeze drying samples.** The cross section of fibers is almost circular which is estimated by measuring the height of individually dispersed fibers.

**Table 1 Molecular stacking distance [Å] (top) and the full width at half maximum (FWHM) [Å$^{-1}$] (bottom) obtained from GIWAXS measurements.**

|  |  | Isotropic phase | Nematic tactoids | Twist-bent mesophase I | Twist-bent mesophase II |
|---|---|---|---|---|---|
| π-π stacking | edge-on | 3.65<br>0.084 | 3.63<br>0.092 | 3.63<br>0.105 | 3.59<br>0.094 |
|  | face-on | 3.61<br>0.113 | 3.61<br>0.106 | 3.59<br>0.099 | 3.56<br>0.103 |
| Lamellar stacking* | edge-on | 25.85<br>0.043 | 25.85<br>0.042 | 25.85<br>0.039 | 25.33<br>0.040 |
|  | face-on | 24.25<br>0.035 | 24.63<br>0.032 | 24.25<br>0.035 | 24.63<br>0.030 |

*Inconsistency of lamella stacking distance with SAXS is possibly due to the fact that solvation has changed the lamella stacking distance.

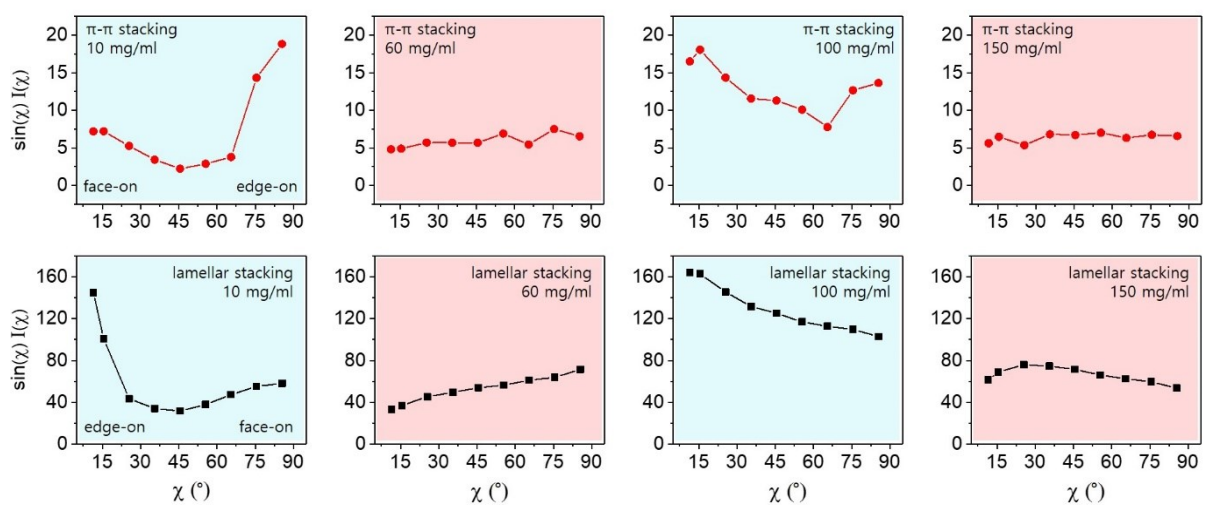

**Figure S4. Geometrically corrected intensity of π-π stacking (010) peak as a function of polar angle, χ.** Note that χ = 0° and 90° correspond to each face-on and edge-on orientation for π-π stacking and χ = 0° and 90° indicated each edge-on and face-on orientation for lamellar stacking.

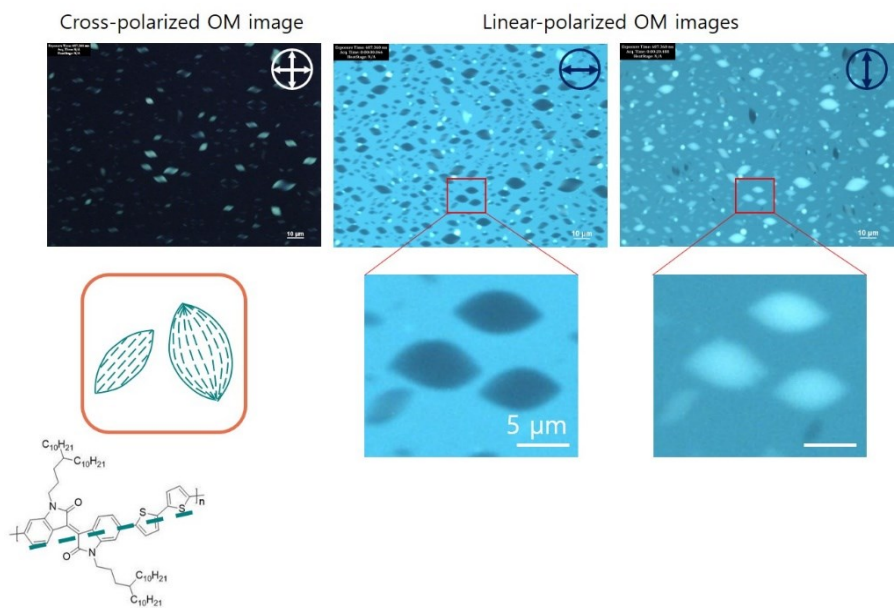

**Figure S5. Cross and linear polarized optical microscopy images of the tactoids.** The tactoids show a uniform change in brightness when the tactoid is rotated from 0◦ to 90◦ relative to the polarizer. The tactoids show uniformly dark and bright when the main axis is aligned parallel and perpendicular to the polarizer, respectively. The scheme (bottom left) shows how the polymer chains are aligned inside the tactoids, indicating the chains are aligned along the fiber long axis.

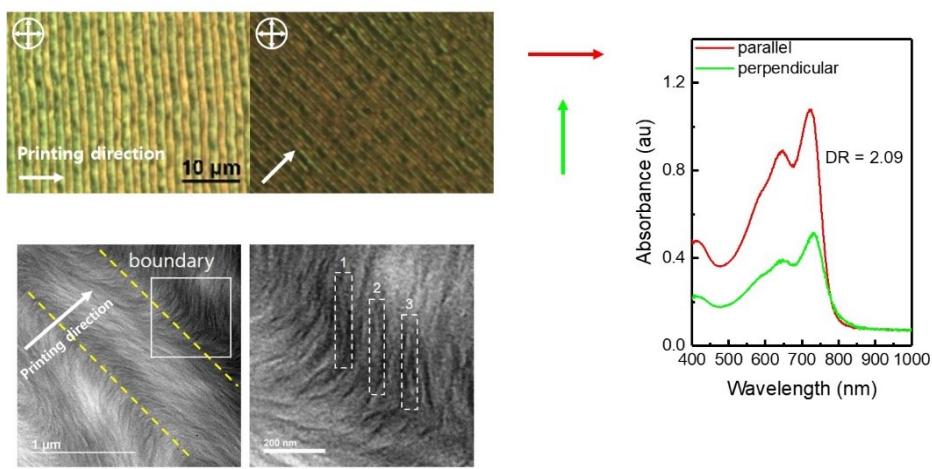

**Figure S6. CPOM, TEM and polarized UV-Vis absorption of printed PII-2T films.** The nanoscale fibers are overall aligned along the printing direction despite they are wavy. Higher optical absorption when the polarizer and printing direction is parallel indicates the polymer chains are aligned along the fiber long axis.

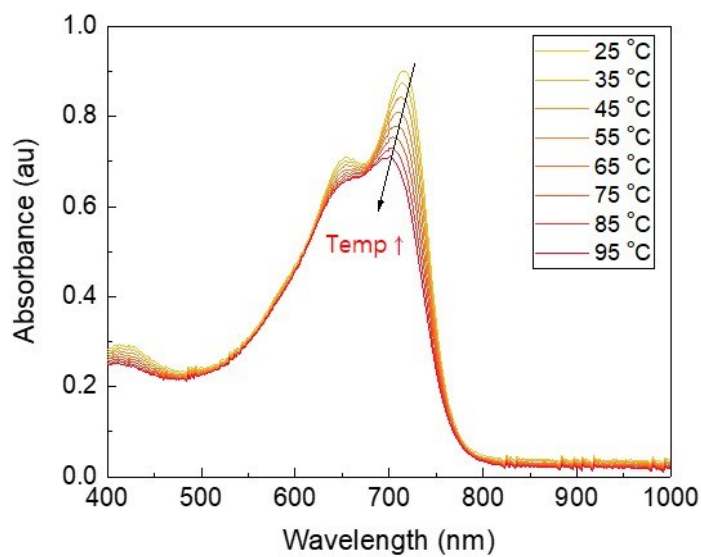

**Figure S7. In-situ thermal UV-Vis absorption of 10 mg/ml PII-2T solution in chlorobenzene.**

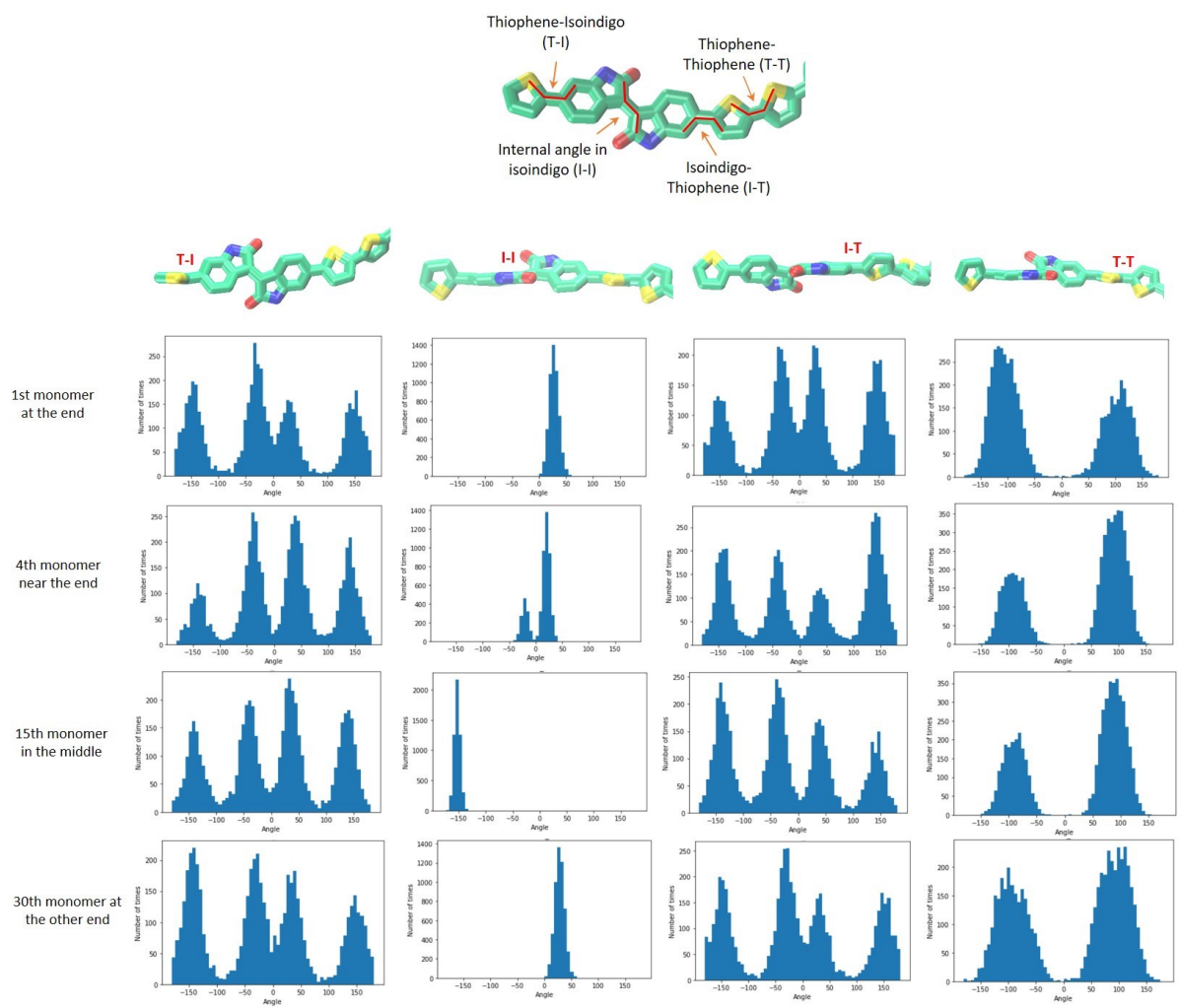

**Figure S8. Dihedral angle change at both ends, near the end and the middle region of the 30-mer PII-2T.** Each angle (T-I, I-I, I-T and TT) is denoted on top of the plots with side view of molecular structures. Most of the dihedral angles are substantially changed; The I-I torsion is the most stable, centered around 25º. The T-I and I-T torsion is highly shaking, centered at two angles of +/–30º and +/–150º. The T-T torsion is also dramatically shaking, centered around +/–100º. The drastic changes of the T-I/I-T and T-T torsion might be the source of dramatic conformational change.

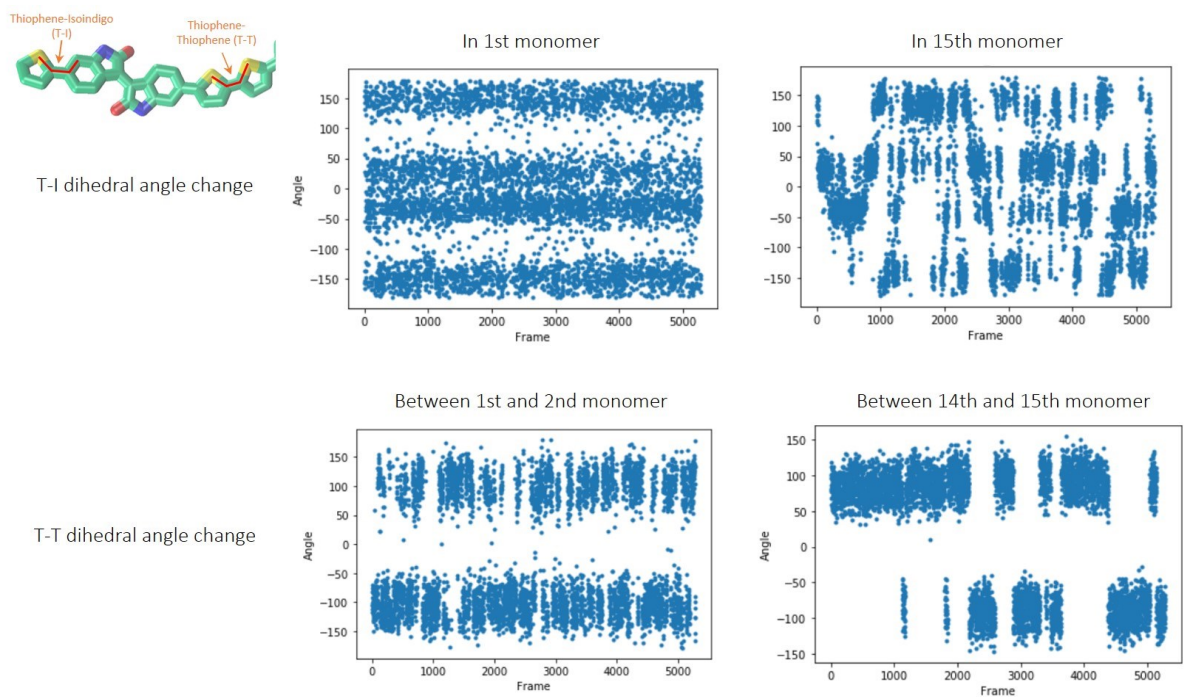

**Figure S9. The comparison of T-I and T-T dihedral angle change between the end and middle region of the 30-mer.** The frequency distribution in a range of $+/-150°$ is similar for both end and middle regions. The distribution in the end region is very much fluctuated whereas the one in the middle region is relatively stable.

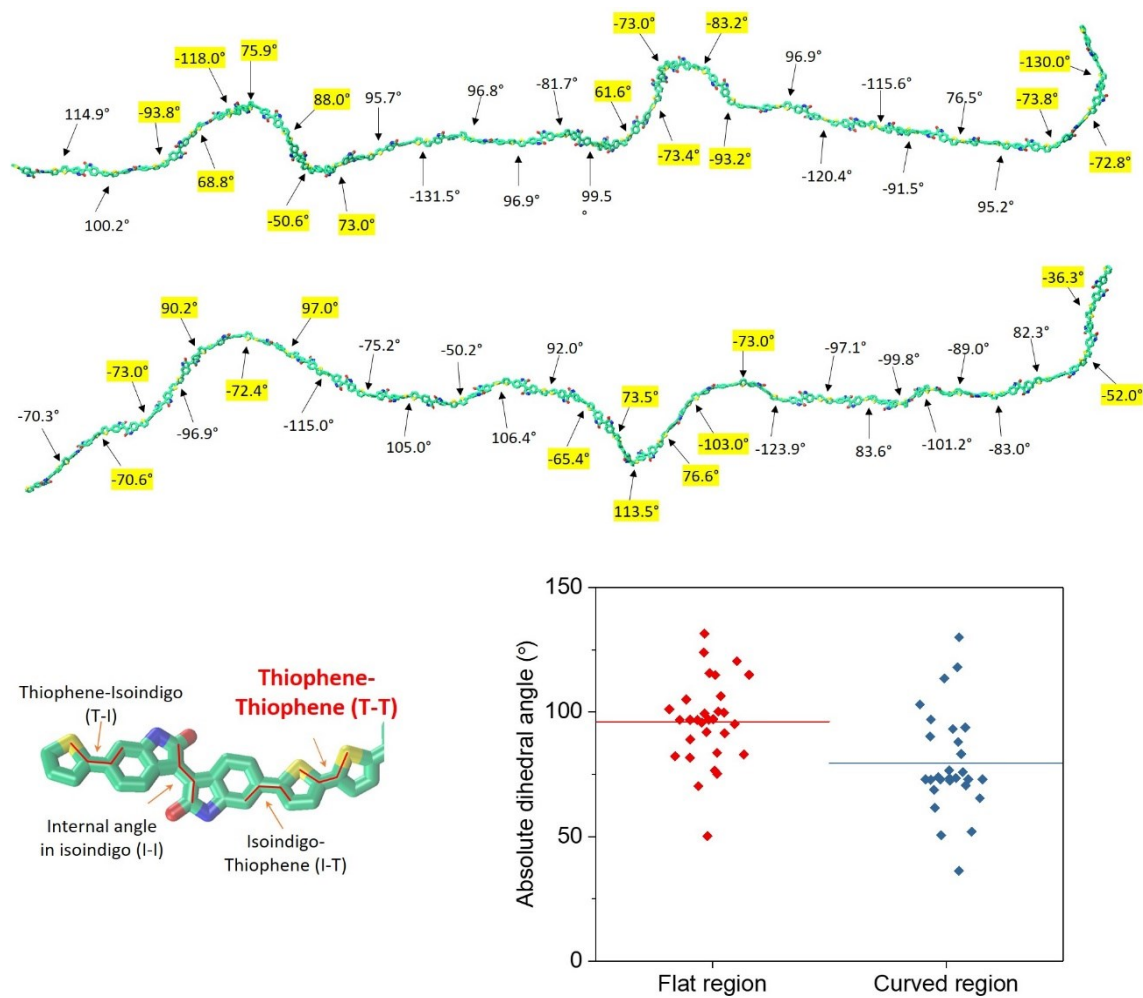

**Figure S10. Selected 30-mer chains captured from the MD simulation (top) and dihedral angle frequency plot from the flat and curved region (bottom, right).** The angles were measured between the thiophene rings (T-T) of each monomer. The flat and curved region is marked by non-highlighting and highlighting, respectively. The plot shows a distribution of angle frequency, with an average value of 96.8 ± 16.9° and 79.4 ± 20.3° for the flat and curved region, respectively.

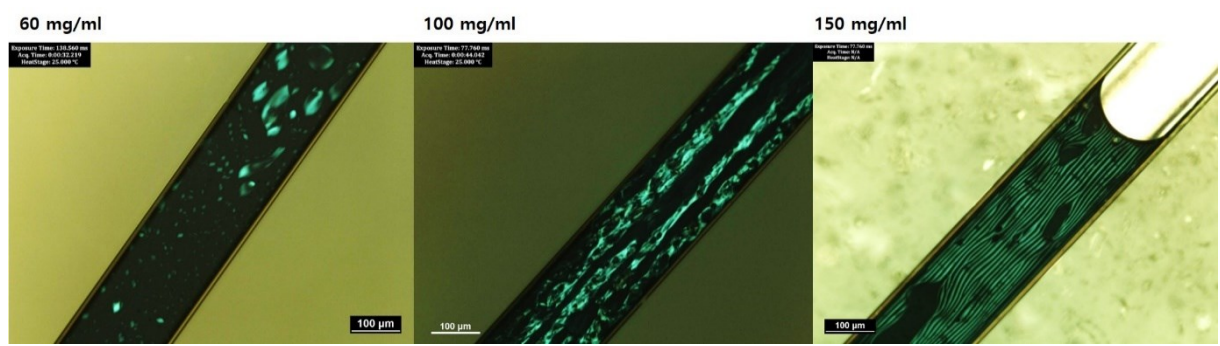

**Figure S11. CPOM images of pristine-made PII-2T solutions in 20-μm-length glass capillary.**

**Movie 1. In-situ cross polarized optical microscopy (CPOM) video of PII-2T solution in a moving, drying meniscus.** The meniscus was created by sandwiching the polymer solution between two glass slides. The video shows the solution phase (dark region on the left top) and the mesophase (bright region in the right bottom). The elliptical mesogenic domains emerge from the solution phase and coalesce to form a rope-like texture.

**Movie 2. Entire trajectory of the 30-mer simulation for 260 ns.** The polymer was put in a box of chloroform. The video shows the flexible conformation of PII-2T chain in solution and the wavy-structure it takes.

**Movie 3. Entire trajectory of two short side-chain PII-2T hexamers simulated for 400ns.** The two molecules were put 20 Å apart in a simulation box solvated by chloroform. This video shows the hexamers took about 100 ns to get aligned and moving together in a stable parallel conformation.